\documentclass{article}
\usepackage{psfig}
\addtocounter{page}{-1}%
\begin{document}

\title{\vspace{-3cm}
\LARGE\bf Comments on the Riemann conjecture and index theory
on Cantorian fractal space-time}
\author{Carlos Castro$^1$ and Jorge Mahecha$^2$ \\
{\small\em $^1$Center for Theoretical Studies of Physical Systems}\\
{\small\em Clark Atlanta University, Atlanta, Georgia, USA}\\
\smallskip
{\small\em $^2$Departamento de F\'{\i}sica, Universidad de Antioquia,
Medell\'{\i}n, Colombia}}
\date{\today}
\maketitle

\begin{abstract}

An heuristic proof of the Riemman conjecture is proposed.
It is based on the old idea of Polya-Hilbert. A
discrete/fractal derivative self adjoint operator whose
spectrum may contain the nontrivial zeroes of the zeta
function is presented. To substantiate this heuristic
proposal we show using generalized index-theory arguments,
corresponding to the (fractal) spectral dimensions of
fractal branes living in Cantorian-fractal space-time, how
the required $negative$ traces associated with those
derivative operators naturally agree with the zeta function
evaluated at the spectral dimensions. The $\zeta (0) = -
1/2$ plays a fundamental role. Final remarks on the recent
developments in the proof of the Riemann conjecture are
made.

\end{abstract}

\section{\bf Introduction}
\label{sec:intro}

Riemann's outstanding conjecture that the non-trivial
complex zeroes of the zeta function $\zeta(s)$ must be of
the form $s = 1/2 \pm i\nu;~ \nu > 0$, remains one of the
open problems in pure mathematics. Starting from an
heuristic study of the index theorem associated with the
dynamics of fractal $p$-branes living in Cantorian-fractal
space-time ${\cal E}^{(\infty)}$ \cite{naschie} we found
some suggestive relations with the Riemann conjecture.

The construction of Cantorian-fractal space-time
\cite{naschie}, ${\cal E}^{(\infty)}$, contains an infinite
number of sets ${\cal E}^{(i)}$, where the index $i$ ranges
from $-\infty, + \infty$. Such index labels the topological
dimension of the smooth space into which the fractal set is
packed densely. For example, the sand on the beach looks
two-dimensional on the surface. This is due to a
coarse-grain averaging/smoothing of the underlying
$3D$-grains which comprise it. In a similar vein the
Hausdorff dimensions of the fractal sets packed densely
inside the smooth manifold of {\bf integer} dimension can
be {\bf larger} than the actual topological dimension of
the space into which is being packed.

The best representative of this is the random backbone
Cantor set, ${\cal E}^{(0)}$, a fractal dust which is
packed densely into a set of topological dimension zero (a
point), and whose Hausdorff dimension equals to the
golden-mean $\phi > 0$, with probability {\bf one},
according to the Mauldin-Williams theorem \cite{mauldin}.
We set the golden mean to be $1/(1+\phi) = \phi = (\sqrt 5
- 1)/2 = 0.618...$. Notice that our conventions differ from
those by Connes in his book \cite{connes}. He chooses for
$\phi = (\sqrt 5 + 1)/2$. We hope this will not cause
confusion.

Incidentally we noted that
\begin{equation}
\phi^k=(-1)^kF_{k-1}+(-1)^{k+1}F_k\phi,~~
\phi^{-k}=F_{k+1}+F_k\phi,~~k=0,\pm1,\pm2,... 
\end{equation}
where $F_k$ are Fibonacci numbers. In this way, ...
$\phi^{-4} = 5 + 3 \phi$, 
$\phi^{-3} = 3 + 2\phi = 4 + \phi^3$, 
$\phi^{-2} = 2 + \phi$, 
$\phi^{-1} = 1 + \phi$, 
$\phi^2 = 1 - \phi$,
$\phi^3 = -1 + 2\phi$,
$\phi^4 = 2 - 3\phi$,
$\phi^5 = -3 + 5\phi$,
$\phi^6 = 5 - 8 \phi$,
$\phi^7 = -8 + 13 \phi$, ... 

The negative values of the topological dimensions signify
the degree of ``emptiness'' or voids inside ${\cal
E}^{(\infty)}$. The simplest analog of this is Dirac's
theory of holes to explain the negative energy solutions to
his equations (positrons/antimatter). Negative entropies
and negative dimensions \cite{nasccast} were of crucial
importance to have a rigorous derivation of why the average
dimension of the world (today) is very close to: $4 +
\phi^3 = 4.236...$.

Negative probabilities and the non-commutative properties
of ${\cal E}^{ (\infty)}$ were essential to explain the
wave-particle duality of an indivisible quantum particle
traversing the Young's double-slit \cite{nasccast}. To be
precise, with the nonlocality of QM which must not be
confused with interference. The non-commutative geometry of
the von Neumann's type associated with Cantorian-fractal
space-time ${\cal E}^{(\infty)}$ is the appropriate
geometry to formulate the new relativity theory
\cite{castro} that has derived the string uncertainty
relations, and the $p$-branes generalizations, from first
fundamental principles (See \cite{ccastro} and
{ccccastro}). Moreover, such new scale relativity, an
extension of Nottale's original scale relativity
\cite{nottale}, is devoid of EPR paradoxes
\cite{alex} and it explains the origins of the holographic
principle \cite{granik}. 

The zeta function has relation with the number of prime
numbers less than a given quantity and the zeroes of zeta
are deeply connected with the distribution of primes
\cite{riemann}. The spectral properties of the zeroes are
associated with the random statistical fluctuations of the
energy levels (quantum chaos) of a classical chaotic system
\cite{main}. Montgomery \cite{montgomery} has shown that
the two-level correlation function of the distribution of
the zeroes is the same expression obtained by Dyson using
Random Matrices techniques corresponding to a Gaussian
Unitary Ensemble.

String theory can be reformulated as a statistical field
theory on random surfaces. For a deep connection between
fractal strings, number theory and the zeroes of zeta see
the book \cite{lapidus}. Recently one of the authors
\cite{cccastro} has been able to establish the link among
$p$-adic stochastic dynamics, supersymmetry and the Riemann
conjecture by constructing the operator that yields the
zeroes of the zeta. The supersymmetric quantum mechanical
model (SUSY QM) associated with the $p$-adic stochastic
dynamics of a test particle undergoing a Brownian random
walk was constructed \cite{cccastro}. The zig-zagging
occurs after collisions with an infinite array of
scattering centers that fluctuate randomly. One can
reformulate such physical system as the scattering of the
test particle about the infinite location of the prime
numbers. Such $p$-adic stochastic process has an underlying
hidden Parisi-Sourles supersymmetry and can be modelled by
the scattering of the test particle by a ``gas'' of $p$-adic
harmonic oscillators--quanta--whose fundamental frequencies
(imaginary) are given by $\omega=i\log p$ and whose
harmonics are $\omega_{p,n}=i\log p^n$. For further details
we refer to \cite{cccastro}. Based on this work Pitk\"anen
was able to give new hints for the proof of the Riemann
conjecture \cite{pitkanen}.

Generalizations of the zeta function exist. Weil has
formulated zeta function of general algebric varieties over
finite fields \cite{weil}. For numerical calculations of
the zeroes, A. Odlyzko has computed up to $10^{22}$ zeroes
of zeta \cite{odlyzko} in agreement with the Riemann
hypothesis.

On a closing note, we honestly feel that theoretical
physics in the new century may dwell on the following
partial list: The new relativity theory $\leftrightarrow$
fractal $p$-branes in Cantorian space-time
$\leftrightarrow$ irrational conformal field theory
$\leftrightarrow$ number theory $\leftrightarrow$
non-commutative (non-associative) geometry
$\leftrightarrow$ quantum chaos (quantum computing)
$\leftrightarrow$ quantum groups $\leftrightarrow$ $p$-adic
quantum mechanics.

In the first part of section {\bf 2} we shall briefly
discuss the basic features of Cantorian-fractal space-time
and heuristically postulate the existence of trace formula
linked to the index of a fractal/discrete derivative
operator. A discussion about the distribution of the
imaginary parts of the zeroes of the Riemann's zeta
function, illustrated by two figures, is presented. In
section {\bf 3} we present a rigorous derivation of the
index-theoretic results based on the $\eta$ (See Eq.
[\ref{eq:eta}]) invariant which is related to the spectral
staircase associated with the spectral dimensions of the
infinite number of hierarchical sets living inside the
fractal strings. Final remarks are made about the recent
developments in the proof of the Riemann conjecture.

\section{\bf Quantum chaos and index theory in ${\cal
E}^{(\infty)}$}
\label{sec:quanchao}

Our motivation was sparked originally by the quantum
counterpart of the classical chaos linked to the ``billiard
ball'' moving on hyperbolic surfaces (constant negative
curvature). As is well known to the experts the Selberg
trace formula is essential to count the primitive periodic
orbits of classical dynamical systems. The spectrum of
(minus) Laplace-Beltrami operator on such hyperbolic
surfaces is linked to the zeroes of the Selberg zeta
function.

Knowing the energy eigenstates of the Schr\"odinger
equation allows to locate the location of the nontrivial
zeroes of the Selberg zeta function.  They also have the
form of $s = 1/2 \pm i p_n$ where $E_n = p^2_n + 1/4$.

One of the most important features of the fractal/discrete
operator is that it has {\bf negative index} as we shall
intend to show. This is just a result of the negative
dimensions/holes/voids of ${\cal E}^{(\infty)}$.  See
\cite{bohm}.

These voids behave like absorption lines in the spectra of
the Hamiltonian associated with the fractal operator ${\cal
D}_f$. Connes already gave a detailed analysis of the
necessity for the trace to be {\bf negative} (absorption
lines) to account for the zeroes of the zeta function
\cite{aconnes}.

The old idea of Polya-Hilbert, is for example, the search
for an equation of the type Schr\"odinger equation on a
hyperbolic space (constant negative curvature):
\begin{equation}
-{\cal D}_f ( {\cal D}_f -1 ) \Psi = s(1-s)\Psi = s {\bar s}\Psi = E_n
\Psi \Rightarrow s = {1\over 2}+ i p_n = {1\over 2}+ i \sqrt { E_n -
{1\over 4 }}. 
\end{equation}

The zeroes of the zeta function are then linked to the
energy eigenvalues $E_n$ in such a way that $\zeta ( 1/2
\pm i p_n ) = 0$ and therefore the zeroes of zeta must lie
in the critical line: $\Re(s) = 1/2$ and the Riemann
conjecture could be proven, at least heuristically.

Quantum groups emerged as a result of inverse scattering
methods. The search is to find now whether such operators
can be constructed. If they can, then their spectrum will
pick up the imaginary part of the zeroes of the zeta.

The sought-after self-adjoint properties of the fractal
derivative, with respect a suitable inner product, are:
\begin{equation}
{\cal D}_f^+ = {\cal D}_f -1.\quad ({\cal D}_f -1)^+=
{\cal D}_f.\quad [{\cal D}_f ({\cal D}_f -1 )]^+ = ({\cal
D}_f -1)^+ {\cal D}_f^+ = {\cal D}_f({\cal D}_f -1).
\label{eq:beltrami}
\end{equation}

Then the Laplace-Beltrami operator is self-adjoint and
$-{\cal D}_f({\cal D}_f -1)$ has a positive definite energy
spectrum. If this fractal/discrete derivative operator
satisfies the properties above, and the operator is
trace-class, then the Riemann conjecture could be proven
following the arguments of Polya and Hilbert.

In general the Riemann-Roch theorem corresponding to a
Riemann surface of genus $g$ is associated with a family of
derivative operators $\nabla^{(n)}_z$. The former is the
derivative operator of ``conformal'' $U(1)$ weight $n$
acting on the family of tensors $T^{(n)}$ with $q$
holomorphic indices and $p$ antiholomorphic indices such
that $q - p = n$.

The index of the operator $\nabla^{(n)}_z$ is defined as
the (complex) dimension of its kernel minus the (complex)
dimension of its cokernel and is equal to $-(n-{1\over2})$
times the Euler number of the Riemann surface of genus $g$
which is given by $2 - 2g$.

In particular, when $n=0$ then $\nabla^{(0)}_z =
\partial_z$ and the index of $\partial_z$ is defined as the
(complex) dimension of the kernel of $\partial_z $ minus
the (complex) dimension of the cokernel of $\partial_z$.
The Riemann-Roch theorem becomes then for $n = 0$
\begin{equation}
Index [\partial_z] = {1\over2} Euler~number = {1\over2} (2 - 2 g) = 1 - g,
\end{equation}
the reason is that the complex dimension is 1/2 the real
dimension, so the alternating sum of Betti numbers
multiplied by an over all factor of 1/2 will select the
complex dimension in compliance with the
Hirzebruch-Riemann-Roch index theorem. The index in this
case depends on the genus of the surface: $g = 0$
corresponds to a sphere, $g = 1$ to a torus and so forth.
For details we refer to Nakahara's book
\cite{nakahara}. 

We are generalizing these results to the case when the
Euler number is given by the alternative sums of Betti
numbers, alternative sums of all dimensions of the possible
cycles. In the case of a two dimensional surface one has
three terms, and only three terms, in the sums only: Euler
number of a two dimensional Riemann surface $= 1 - 2g + 1 =
2 - 2g$.

We will show that the index corresponding to the
fractal/discrete derivative operator on the fractal world
sheet ${\cal E}^{(2)}$, whose fractal dimension is $1 +
\phi$, is given by the value of the zeta function evaluated
on the spectral dimension of the infinite-cycle
intersection $E_{infinity}$ which equals $dim~E_{infinity}
= (s) = 0$.

Spinors on Riemann surfaces are defined in terms of the
square roots of the $n=1$ line bundles: The weight $1/2$
corresponds to positive chirality spinors and the weight $-
1/2$ corresponds to negative chirality ones.

In Cantorian fractal space-time we may follow by analogy
similar arguments if one takes into account that the
intrinsic fractal dimension of a bosonic random walk is $1
+ \phi$. This means that dimensions are counted in basic
units of the latter dimension. In particular, the intrinsic
dimension of a fermionic random walk is $1/2$ the bosonic
one: $(1/2)(1+\phi)$.

The analog of the higher genus Riemann surfaces corresponds
to spaces of negative dimensions. In Cantorian-fractal
space-time, the totally void set, ${\cal E}^{(-\infty)}$,
is the one whose fractal dimension is equal to zero and is
embedded in a space of $- \infty$ topological dimension.
The index associated with the analog of the $n = 0$
derivative operator $\partial_z$ in ordinary Riemann
surfaces of genus $g$, in Cantorian-fractal space-time is
no longer an integer!, and is defined in basic units of $1
+ \phi$. It is evaluated on the infinite-cycle-intersection
$E_{infinity}$ space and is given by the analog of the
Riemann-Roch theorem:
\begin{eqnarray}
``Index" {\cal D}&=&``Trace" [{\cal D}^{-(s) }] = -(0 -
\frac{1}{2})(1 + \phi)~Euler~[E_{infinity}] \nonumber\\
&=&{1\over2}(1 + \phi)(-\phi) = -{1\over2} = \zeta(0)
\end{eqnarray}
and this agrees exactly with the value of $\zeta(0)$ since
the dimension of the infinite-intersection cycle
$E_{infinity}$ is precisely $(s) = 0$:
\begin{equation}
dim~E_{infinity} = (s) = 1\cdot\phi\cdot\phi^2\cdot\phi^3\cdot\,
...\,\phi^{s - 1}
\end{equation}
is $0$ in the $s =\infty$ limit.

Therefore, using fractal derivatives and/or discrete
derivatives like they occur in quantum-groups,
$q$-calculus, and in $p$-adic QM, and studying the spectrum
of fractal strings/branes in ${\cal E}^{\infty}$
space-times one may try to use the analog of the
Riemann-Roch theorem:
\begin{equation}
``Index" [{\cal D}_{fractal}] (E_{(s)}) = ``Trace" [{\cal
D}_{fractal}^{-(s)}] ={\zeta (s)} \sim {Euler} [E_{(s)}],
\label{eq:RiRoanalog}
\end{equation}
where the subspace $E_{(s)}$ (where the index is restricted
on) of the world sheet ${\cal E}^{(2)}$ is some suitable
{\bf intersection} of a collection of sets, or cycles of
${\cal E}^{(\infty)}$ living {\bf inside} the world sheet
${\cal E}^{(2)}$.
\begin{equation}
{\cal E}^{1} \wedge {\cal E}^{0} \wedge {\cal E}^{-1}... \wedge {\cal
E}^{-s + 1},
\label{eq:intersecc}
\end{equation}
whose dimension is
\begin{equation}
dim~E_{(s)} =1\cdot\phi\cdot\phi^2\dot...\phi^{s-1} = \phi^{s(s-1)/2},
\label{eq:dimension}
\end{equation}
where $(s)\equiv s(s-1)/2$ and $s$ counts the number of
cycles involved in the intersections, and whose Euler
number is given by the usual formulae (alternating sums of
Betti numbers):
\begin{equation}
{Euler} [E_{(s)}] = \sum_{k=-s}^1 (-1)^k\phi^{- k+1} =
\frac{-1+(-1)^s\phi^{s+2}}{1+\phi} = \frac{-1+F_{1+s}-\phi
F_{2+s}}{1+\phi}. 
\label{eq:alternating}
\end{equation}
One must not confuse the ${\cal D}_{fractal}$ in the trace
operation with the ${\cal D}_f$ of eq. (\ref{eq:beltrami}).

The operator (Dirac) ${\cal D}_{fractal}$ is related to the
Laplace operator by ${\cal D}_{fractal}^2$ = Laplace, used
in the construction of tbe spectrum of fractal strings by
Lapidus et. al. \cite{lapidus}. These authors have shown a
relation among the sequence of frequencies and lengths of a
fractal string with the eigenvalues $\lambda$ of the
Laplace operator (zeta function).

The geometric counting function $Z_L(s)$, the frequency
counting function $Z_F(s)$ and the zeta function
$\zeta(s)$, for $s$ = complex dimension, are related by
\begin{equation}
Z_F(s) = \zeta_{gen}(s) Z_L(s),
\end{equation}
where the generalized $\zeta(s)$ is
\begin{equation}
\zeta_{gen}(s) = \sum \lambda^{-s/2},
\end{equation}
with $\lambda$ the eigenvalues of the Laplacian. In the
case of Bernoulli string $\lambda\sim n^2$ so $\zeta_{gen}$
coincides with the Riemann zeta. For fractal drums one gets
the Epstein zeta function. See reference \cite{lapidus} for
details.

Equation (\ref{eq:alternating}) is the explicit expression
of the generalized Euler number as an alternative sum of
the dimensions of the higher-dimensional voids/holes
(``genus'' of the fractal string). Notice that in the
asymptotic limit the alternating series converges exactly
to: ${Euler} [E_{(\infty)}] = - \phi <0$! which is a clear
indication that Cantorian-fractal space-time is {\bf
left-handed}. This asymmetry between right/left chirality
is also very natural in Penrose's twistor theory.

Higher dimensional (than one) sets: dim = $(1+\phi)^k$
correspond to fractal $p$-branes, are those corresponding
to the values of $p = (1+\phi)^k > 1$. The sets of negative
topological dimension are higher-dimensional $holes/voids$:
They play the role of the $higher~ genus$ surfaces in
Cantorian-fractal space-time. The backbone set ${\cal
E}^{0}$ will play the role of the discrete/fractal flow of
time associated with the fractal string which is
represented by an open subset of the normal set ${\cal
E}^{1}$ whose dimension is equal to $1$. Such
discrete/fractal temporal evolution of the fractal string
has for fractal world-sheet, ${\cal E}^{(2)}$ of fractal
dimension $1+\phi$, given by direct product of the backbone
set ${\cal E}^{0}$ times ${\cal E }^{(1)}$ and whose
higher-dimensional voids/holes or higher ``genus Riemannian
surfaces'' are nothing but the rest of the sets of negative
topological dimensions.

In (super) strings, the multi-loop scattering amplitudes
depend crucially on the suitable integrals over the (super)
moduli space of the higher-genus surfaces. The Selberg zeta
function plays an essential role in providing proper
counting of the number of the primitive closed geodesics
that tasselate the hyperbolic space, providing with a
single cover of the (super) moduli space. This occurs for
genus higher than $1$ (the torus).

The parameter $k$ which defines the lower bound of the
alternating sum for the Euler number is nothing but the
{\bf analog} of the ``genus'' of the world-sheet associated
with this {\bf fractal ``string''} ``living'' in ${\cal
E}^{\infty}$. A $p$-brane spans a $p+1$ world volume
generated by its motion in time which accounts for the
extra dimension: A string spans a 2-dim world-sheet; a
membrane a 3-volume and so forth. All this is naturally
related to the statistical properties of random matrix
models in lower dimensions in the large $N$ limit;
irrational conformal field theories; irrational values of
the central charges; the monster group, etc.

From an elementary numerical calculation of equation
(\ref{eq:alternating}), we can see that the Euler number of
the multiple cycle-intersection $E_{(s)}$, as a function of
the powers of $\phi^k$ (the genus-analog) {\bf oscillates}
about the golden mean. It is well known that the
distribution of primes oscillates abruptly like the
spectral-staircase levels in quantum chaos.

In the limit of infinite ``genus'', the Euler number, the
alternating oscillatory sum will converge to the golden
mean with a {\bf negative sign}, consistent with the nature
of the absorption lines linked to the holes/voids/genus of
the world sheet of the fractal string. As remarked earlier,
Connes emphasized the importance that a {\bf negative}
value of the index theorem in non-commutative geometry must
have to understand the location of the zeroes of zeta
function as absorption spectral lines.

To sum up what has been said so far: Using
$fractal/discrete$ derivatives ${\cal D}_f$ which are the
more appropriate ones for ${\cal E}^{(\infty)}$ we make
contact with the ``Riemann zeta'' function associated with
such Cantorian-fractal space-time. As said previously,
discrete derivatives are very natural in the $q$-calculus
used in quantum groups (Jackson's calculus) and there is a
deep relation between $p$-adic quantum mechanics and
quantum groups as well.

The initial reasons why we believe a trace formula may be
valid in Cantorian-fractal space-time is the following
argument, despite the fact that we don't get a perfect
matching of numbers. But they are close.

Looking at equation (\ref{eq:alternating}) for the first
entry associated with the {\bf triple} cycle-intersection
of the three sets: ${\cal E}^0; {\cal E}^1; {\cal E}^{-1}$,
we find for the Euler number associated with the
alternating sum of dimensions:
\begin{equation}
{Euler} [E_{(3)}] = -1 + \phi - \phi^2 = -1-1+2\phi=-1+\phi^3= {\phi^3
\over 2} - (1 - {\phi^3 \over 2}). 
\label{eq:euler3}
\end{equation}

The fractal dimension of the intersection of these three sets,
intersection of three {\bf cycles} is: 
\begin{equation}
dim~E_{(3)} = (1) (\phi) (\phi^2) = \phi^3 = 0.236068.
\label{eq:dimE3}
\end{equation}

Now evaluate the ``index'' for this particular case: 
\begin{eqnarray}
``Index" [{\cal D}_{fractal}] (E_{(s)}) &=& ``Trace" [{\cal
D}_{fractal}^{-(s)}] = \zeta (s = \phi^3) = -0.790068\nonumber\\
&\sim&{Euler}[E_{(3)}] = - 0.763932 < 0.
\label{eq:indexE3}
\end{eqnarray}

Notice how close our answer was: $-0.790068 \sim -
0.763932$. Although not a perfect match, this is a good
sign that we are on the right track.  Looking down, one can
see that the numbers do not differ much. The index of the
exterior derivative operator associated with the de Rham
elliptic complex coincides with the Euler number, for
example.

The ``trace'' in non-commutative geometry as Connes has
emphasized many times:
\begin{equation}
``Trace"[{\cal D}_{fractal}^{-(s)}] \leftrightarrow {\rm
volume ~of~ the~ space~ that~ has~ dimension}~ (s).
\label{eq:trace}
\end{equation}
One must be very careful not to confuse the label $``s"$
with the label $``(s)"$, they are not the same. Only in the
special case $E_{(3)}$.

To justify further this proposal, for example, lets take
now a look at the quadruple intersection (for real
dimensions). The quadruple intersection of four cycles:
\begin{equation}
dim~E_{(4)} = 1\cdot \phi\cdot\phi^2\cdot\phi^3 = \phi^6 =
\phi^{4(4-1)/2} = 0.0557281.
\label{eq:4-intersecc}
\end{equation}

So evaluating the Euler number from the alternating
sum/Betti numbers from k = -2,-1, 0, 1:
\begin{eqnarray}
&&``Index"~[{\cal D}_{fractal}](E_{(4)}) = ``Trace"[{\cal D}^{-(s)}] =
{\zeta (\phi^6)} = - 0.55451\nonumber\\
&\sim& Euler~[E_{(4)}] = 2\phi^3 - 1 = \phi^3 - (1 - \phi^3) = -0.527864 <
0. 
\label{eq:eulerE4}
\end{eqnarray}
Notice once again that the numbers are not so far off!,
\begin{equation}
- 0.55451 \sim -0.527864.
\end{equation}

If one looks at the asymptotic infinite-dimensional
voids/holes (``genus'') limit, to extract non-perturbative
information, when the number of intersections of the sets
of negative dimensions goes to infinity, the Euler number
(for real dimensions) converges to the golden mean.
Therefore in the asymptotic limit we have by looking at the
last entries of our tables and at the limit of formula
(\ref{eq:alternating}):
\begin{eqnarray}
{\zeta (0)} &=& - {1\over 2},{\rm~~but}\nonumber\\
Euler~[E_{(infinity)}] &=& -\phi = - 0.618033.
\end{eqnarray}

In this limit one can see that the space clearly is
left/right chiral {\bf asymmetric} like twistors! The
number of self dual modes is not equal to the number of
anti-self-dual ones ...

\begin{figure}[h]
\centerline{\psfig{figure=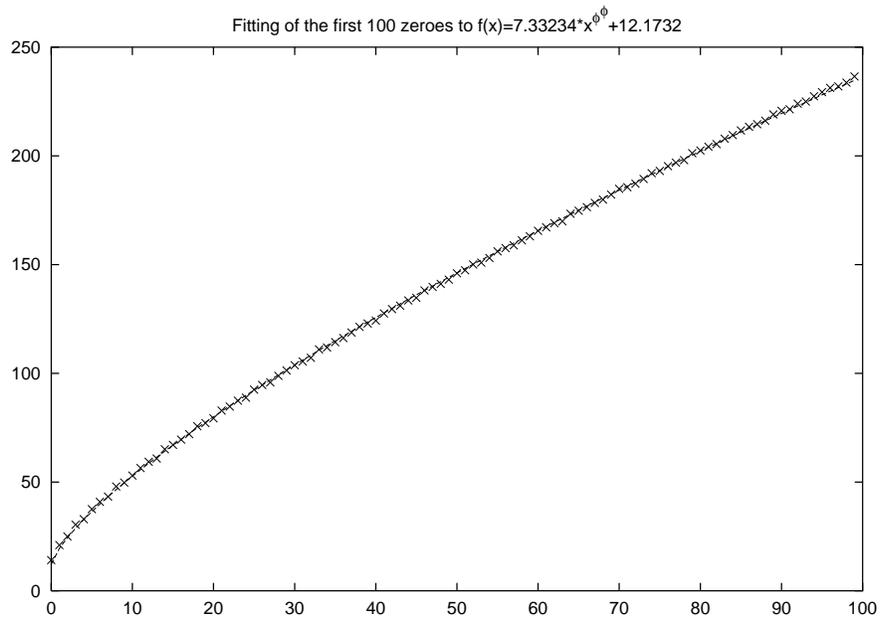,width=12cm,angle=270}}
\vspace{0.0mm}
\caption[Short Title]{Distribution of the imaginary parts
of the zeroes of zeta as a function of the integers
$n=1,2,3,...$. This plot displays the fit $y_n=A
n^{\phi^\phi}+B$; $1\le n\le100$. The crosses are data
obtained by Odlyzko \cite{odlyzko}, and the continous line
is the fitting function. For $A$ = 7.33234 and $B$
= 12.1732 the fit is almost perfect.}
\label{fig:hundred}
\end{figure}

\begin{figure}[h]
\centerline{\psfig{figure=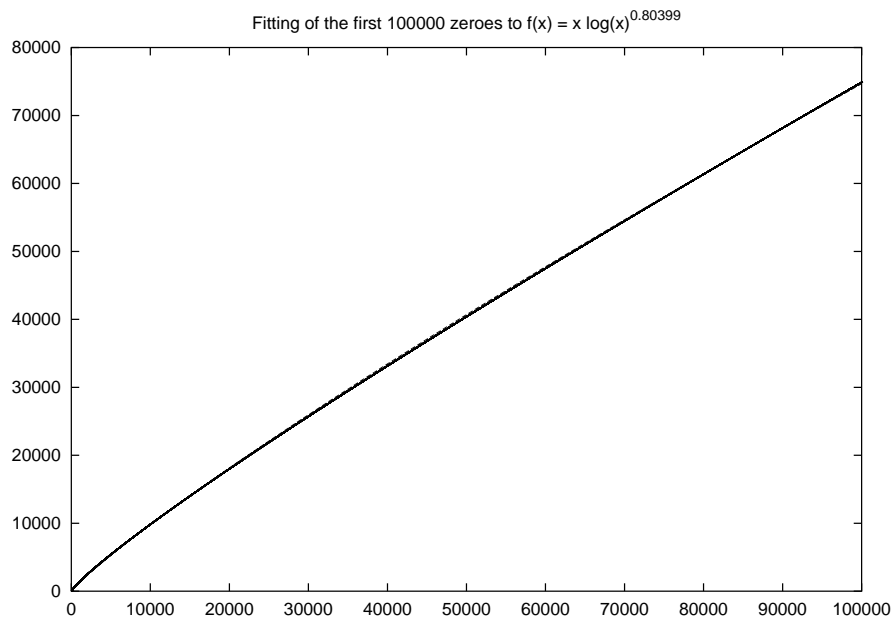,width=12cm,angle=270}}
\vspace{0.0mm}
\caption[Short Title]{Distribution of the imaginary parts
of the zeroes of zeta as a function of the integers
$n=1,2,3,...$. This plot displays the fit $y_n=
(n\log{n})^A$; $1\le n\le10^5$. It is based on arithmetic
QFT estimates where $y_n\approx(n\log{n})^{(1+\phi)/2}$.
The Hamiltonian of arithmetic QFT is $H=\sum_p p\log{p}$;
$p$ = prime. Notice that $(1+\phi)/2\approx 0.80399$. The
dots are data obtained by Odlyzko \cite{odlyzko}, and
the continous line is the fitting function.}
\label{fig:thousand}
\end{figure}

Figure \ref{fig:hundred} shows a suggestive relation
between the zeta function and the golden mean observed in a
fitting of the first 100 zeroes. Figure \ref{fig:thousand}
shows a nice fitting inspired in the Hamiltonian of
arithmetic QFT.

The quantum field theory associated with the geometrical
excitations of Cantorian-fractal space-time is related to a
Braided-Hopf-quantum-Clifford algebra: Braided statistics,
etc... The Clifford-lines in {\bf C}-space (Clifford
spaces) are the Clifford-algebra valued (hyper-complex
number-valued) lines which are the extensions of Penrose
twistor based on complex numbers.

In this asymptotic limit the departure between the $\zeta
(dim~E_{(\infty)})$ and the Euler number is the greatest. 

How do we reconcile the fact that the $\zeta (0) = -{1\over
2}$ differs from the Euler number of $E_{infinity}$ given
by $-\phi$? This is where we invoke the analog of the
Riemann-Roch theorem associated to the derivative operator
$\partial_z$, where now one has holomorphic/antiholomorphic
differentials of fractional weight, in units of the
intrinsic fractal dimension of a fractal world-sheet which
coincides precisely with the dimensions of a bosonic random
walk so that:
\begin{eqnarray}
Index &=& Trace [{\cal D}^{-(s)}](E_{infinity}) = \zeta (0)
= - { 1\over 2}
\nonumber\\
&=&- ( 0 - {1\over2}) ( 1+\phi) Euler~[E_{infinity}] = {1\over2} ( 1+\phi) 
(-\phi) = - {1\over2}.
\end{eqnarray}

Roughly speaking, we are taking the infinite cycle
intersections in the Grassmanian space that encodes the
``Riemann surfaces'' of arbitrary ``genus''. Each ``point''
of the ordinary Grassmanian represents a Riemann surface of
a given genus. Cantorian-fractals space-time is endowed
with a $p$-adic topology where every disc is either
contained inside another disc or it is disjoint from the
latter. Every point inside a disc is a center, meaning that
the center of a disc can occupy two different places
simultaneously.

Similarly one could define spinors as the square root of
the line bundles, $n = 1$ in units of $1+\phi$ so the spin
bundles will be characterized by the following values of
conformal weights: ${|\zeta (0)|/ \phi} = |\zeta (0)|(
1+\phi) = 1/2(1+\phi)$. As strangely as it may seem,
fractional spin and fractal statistics is something which
is not so farfetched. For references see \cite{sidharth}
and \cite{cruz}. Fractional charges, for example, are
natural ingredients in the quantum Hall effect.

The analog of the Riemann-Roch index theorem applied to
this operator would be proportional to the Euler number, in
the infinite ``genus'' case, in the infinite
cycle-intersections, if this operator corresponds to the
conformal weight $n = 0$ derivative operator $\partial_z$,
in units of the intrinsic fractal dimension of a bosonic
random walk, $1+\phi$:
\begin{equation}
Index = \zeta (0) = - (0 - {1\over2})(1+\phi)(- \phi) = - {1\over2}
\end{equation}

This index theoretic part of the paper is the main result.
Besides, something very interesting occurs. The intrinsic
dimension of a fractal Brownian walk of a boson is $1+\phi$
(and the fractal world-sheet as well). The dimension of a
fermion random walk is $1/2$ (dimension of a bosonic random
walk): $(1 + \phi)/2$. This is precisely what we get. The
extrinsic/embedding fractal dimension of a bosonic random
walk is $2$. The extrinsic/embedding dimension of the
fermionic random walk is $(1/2)(2)= 1$.

In ordinary Conformal Field Theory, the central charge of a
boson is $1$ which is equal to the topological dimension of
a path. The central charge of a fermion is $1/2$.

What about spin statistics when we have fractal spin
dimension? The dimension appear in units of $1+\phi$. One
has then an effective Planck constant $\hbar$ (See
\cite{castro}) that will be then:
\begin{equation}
\hbar_{eff} =(1+\phi)\hbar. 
\end{equation}
So a fermion, for example, will have spin = $1/2$ in units
of $\hbar_{eff}$ instead of in units of the usual $\hbar$.

Notice how crucial is the fact that $\zeta(0) = - 1/2$ in
all these results. So essentially, the index theoretic
results only make sense in the critical strip where real
part of $s$ lies between $0$ and $1$.

We need to examine the validity of the Riemann-Roch theorem
for fractal Riemann surfaces and for higher dimensional
surfaces as well, etc ... It is all heuristic but it may
lead to a plausible clue which may shed some light in
proving the Riemann conjecture.

What we shall explore the opposite scenario: What if there
are non-trivial zeroes violating the Riemann conjecture?

We get approximate numbers, when the number of cycle
intersections is finite, but not exact. This has a simple
explanation. The zeta function for real values of $s$ lying
between $0,1$ with $s = \phi, \phi^2, \phi^3,...$ is a
slowly decreasing function while the Euler number {\bf
oscillates}.

A natural thing is to evaluate the zeta function at {\bf
complex} dimensions and check whether a matching with the
Euler numbers (for complex dimensions) occurs; i.e to see
if one can extend the analog of the Riemann-Roch theorem.
Before we evaluate the Euler numbers for complex dimensions
it is important to emphasize that there are no zeroes of
zeta (besides the trivial ones $s = - 2N$) in the region
$\Re(s) < 0$. The argument goes as follows, the functional
equation obeyed by the $\zeta (s)$ is of the form (See
\cite{titchmarsh}):
\begin{equation}
\pi^{-s/2}\zeta(s)\Gamma({s\over2}) =
\pi^{-(1-s)/2}\zeta({1-s})\Gamma({{1-s}\over2}).
\label{eq:eulergamma}
\end{equation}

Since the gamma function has trivial poles at $s = - 2 N$,
the $\Gamma(s/2) = \Gamma(-N) = \infty$, this implies that
$\zeta(s)$ will have trivial zeroes when $s = - 2N$. This
is due to the fact that the right hand side of the previous
equation is well defined as functions of $1 - s$. When $s =
0$, $\zeta (0) = - 1/2$ and the pole of the $\Gamma(0) =
\infty$ corresponds precisely to the only pole of $\zeta(1
- 0) = \infty$ in the right hand side of the equation.

Notice that the critical zeroes of the Riemann zeta
function, $s = {1\over 2} + i\nu$, lie {\bf exactly} in the
vertical line between the following vertical lines of
complex dimensions $\phi + i\nu$ and $\phi^2 +
i\nu=1-\phi+i\nu$:
\begin{equation}
{1\over 2} + i\nu = {1\over 2} [(\phi + i\nu) + (1-\phi +
i\nu) ] = s \equiv {\rm critical~zeroes}.
\label{eq:verticalline}
\end{equation}

For plausible violations of the Riemann conjecture, {\bf
inside} the critical strip, it would be interesting to look
at the behavior of the following points which are
symmetrically distributed about the vertical critical line
$\Re(s) = 1/2$:
\begin{equation}
\zeta(\phi + i\nu){\rm~ and~} \zeta(1 - \phi +i\nu) 
\end{equation}
for example, where $\nu = \Im(s)$, imaginary part of the
non trivial zeroes of zeta, $s = 1/2 + i\nu$.

It would be interesting to examine the behavior of the zeta
evaluated at these points for all values of $\nu$, the
imaginary parts of the nontrivial zeroes and/or other
values of $\nu$. And to see what is the behavior of such
values of zeta for large $ \nu $. How fast they approach or
depart from zero, etc...

Let's start now with complex-valued dimensions and evaluate
$\zeta$ there:
\begin{equation}
\zeta (\phi + i\nu).\quad \zeta (\phi^2 + i\nu).\quad \zeta (\phi^3 +
i\nu),...~\zeta (\phi^n + i\nu),... 
\label{eq:zetaincomplex}
\end{equation}

Let us {\bf foliate} this critical strip (between $\phi$
and $\phi^2$) by an infinite family of {\bf horizontal
lines} passing through each single one of the imaginary
parts of the critical zeroes: The horizontal lines at $\pm
i\nu$ will do the job. The critical strip is comprised of
the two vertical lines in the complex-dimension-plane which
are symmetrically distributed with respect the critical
line: $\Re(s) = 1/2$. In reference \cite{lapidus}, where
Lapidus and Frankenhuysen discuss complex dimensions like
$d=\rho+i\sigma$, $\rho$ is related to ``oscillations of
the geometry'' and $\sigma$ to ``oscillations in sound''.

The ``index'' $[{\cal D}_{fractal}] (E_{(s)})$ is evaluated
now for a particular family of {\bf complex} dimensions of
the spaces given by the {\bf triple} cycle intersections of
${\cal E}^{(1)}, {\cal E}^{(0)}, {\cal E}^{(-1) }$:
\begin{equation}
1 \pm i\nu.\quad  \phi \pm i\nu. \quad  \phi^2 \pm i\nu. 
\label{eq:tripleinters}
\end{equation}

Riemann discovered \cite{titchmarsh} that $\zeta(s)$ has an
analytic continuation to the whole complex plane except for
one simple pole at $s=1$ with residue equal to one. These
properties have immediate consequences on the zero
distribution of $\zeta(s)$: There are no zeros in the
half-plane $\Re(s) > 1$, there are the so-called {\bf
trivial} zeros at $s=-2N$ for every positive integer $N$,
but no others zeros in $\Re(s)<0$. Therefore, {\bf
nontrivial} zeros can only occur in the critical strip
$0\leq\Re(s)\leq 1$.

Equation (\ref{eq:tripleinters}) furnishes these hypothetic
complex dimensions where we could test the validity of the
Riemann-Roch index theorem: Located at the complex plane
points
\begin{equation}
(1+i\nu)(\phi + i\nu) (\phi^2 + i\nu),~~\nu > 0,~~\zeta(1/2+i\nu)=0.
\end{equation}
One must also add the complex conjugates.

This triple product of dimension equals to,
\begin{equation}
(-1+2\phi-2\nu^2)+i(2\phi\nu-\nu^3)
\end{equation}

Two comments are in order: 

$\bullet$ The lowest $\nu$ is bigger than 14. Then we
encounter that $(1+i\nu)(\phi + i\nu) (\phi^2 + i\nu)$, for
every $\nu$, has a negative real part. Then, $\zeta$ {\bf
cannot} have zeroes at $(1+i\nu)(\phi + i\nu)(\phi^2 +
i\nu)$.

$\bullet$ Because the $\zeta$ evaluated in the following
regions where $\Re(s) < 0$ is large: $\zeta[(1 + i\nu)(\phi
+ i\nu)(\phi^2 + i\nu)]$ is exponentially very large and
the values of the Euler numbers corresponding to these
complex dimensions are not of that magnitude, one cannot
longer relate the values of the zeta evaluated at complex
dimensions with the Euler number times the fractal
dimension of a bosonic random walk. Hence, the analog of
the Riemann-Roch index theorem seems to be valid only in
the critical strip: $0 < \Re(s) < 1$.

Concluding this section: If, and only if, a
fractal/discrete derivative operator ${\cal D}_{fractal}$
is found to obey the properties described in this section
then the Riemann conjecture could be proven heuristically.

\section{\bf The Atiyah-Patodi-Singer index theorem and
fractal strings}
\label{sec:frastring}

So far our arguments have been heuristic. It is desirable
to have a more rigorous derivation of these results. The
physical picture we are proposing is that of a string (one
dimensional object) with a discrete fractal flow of time.
The discrete/fractal time is now represented by the random
Cantor set $S^0 \equiv{\cal E}^{(0)}$ or fractal dust of
dimension equal to the golden mean $\phi$ embedded in a
space of topological dimension equal to $0$, a ``point''.
The one dimensional string is represented by the normal set
$S^1 \equiv{\cal E}^{(1)}$ where its fractal dimension
equals $1$ and is embedded in space of topological
dimension equal to unity as well.

The world-sheet spanned by this fractal string has the
topology of $S^1\times S^0\equiv{\cal E}^{ (2)}$ and
corresponds to a space of fractal dimension equal to $1 +
\phi$ embedded in a space of topological dimension equal to
$2$. This is what we referred earlier as the
intrinsic/extrinsic dimension of a fractal bosonic random
walk respectively.

Since the flow of time is discrete/fractal we can represent
the topology of $S^1\times S^0$ as that of an ``even''
dimensional manifold (since the fractal world sheet is
embedded in a two-dimensional space) with {\bf boundaries}.
The boundaries are represented precisely by the
discrete/fractal flow of time and the string itself
corresponds to the $odd$ dimensional manifold $ S^1$.

The appropriate index theorem for Dirac operators in
compact manifolds with boundaries is the
Atiyah-Patodi-Singer index theorem (spectral flow) and the
relevant invariant is the so called $\eta$ invariant given
by the spectral asymmetry of the eigenvalues $\lambda_k$ of
the Dirac operator defined in an odd-dimensional manifold.
After a suitable regularisation is made the $\eta$
invariant is:
\begin{equation}
\eta = \sum_ {\lambda_k > 0} 1 - \sum_{\lambda_k <0} 1 =
\sum_{\lambda_k}{}^{'} sgn(\lambda_k)|\lambda_k|^{-2s},
\label{eq:eta}
\end{equation}
for $Re(s)>0$. The prime means that the zero modes have
been omitted.

We will proceed by analogy along similar lines as the
definition of the $\eta$ invariant. The main difference is
that we are concerned solely with the {\bf spectral
dimension} distribution of the infinite hierarchy of Cantor
sets living inside the fractal string ${\cal E}^{(1)}$.
These are the infinite hierarchy of sets: $S^1, S^0,
S^{-1}, S^{-2}, ...  S^{-\infty}$ of fractal dimensions $1,
\phi, \phi^2, ... \phi^n, ...$ embedded in spaces of
topological dimensions $1, 0, -1, -2, ... -n,
... -\infty$ respectively. 

Therefore we will perform the sums over all topological
dimensions less than $1$ and use the standard zeta function
regularisation (analytic continuation) associated with the
$\sum 1=\infty$ summation:
\begin{eqnarray}
\eta[{\cal E}^{(1)}] &=& \sum_{dim >1} 1 - \sum_{dim < 1} 1 = - \sum_{dim
< 1} 1 \nonumber\\
&=&-\left[1+\sum_{d = -1}^{d=-\infty} 1\right]=
-[1+
\zeta (0)] = -1/2=\zeta(0)
\end{eqnarray}

This is the analog of the spectral staircase, where we are
counting the dimensions from $-\infty$ to $0$. This value
of $\zeta(0)$ is precisely what corresponds to the index of
the fractal derivative operator evaluated on the infinite
-intersection-cycle $E_{infinity}$. Such infinite number of
cycles are nothing but the infinite hierarchy of Cantor
sets living inside ${\cal E}^{(1)}$ and whose topological
dimension of their corresponding embedding spaces are $1,
0, -1, -2, ... -\infty$ respectively. The
higher-dimensional voids correspond to $-1, -2, ...
-\infty$.

Therefore:
\begin{eqnarray}
Index({\cal D})[E_{infinity}] &=& Trace({\cal D}^{-(s)})\nonumber\\
&=&\zeta(dim~E_{infinity}) = \zeta(0) = - 1/2 = \eta[{\cal E}^{(1)}].
\end{eqnarray}

Now we can see why the analog of the Riemann-Roch theorem,
associated with an operator of $n = 0$ conformal weight
$\partial_z$, in the Cantorian-fractal world-sheet becomes
then:
\begin{equation}
Index = {1\over2}(1 + \phi)(- \phi) = (dim_C[{\cal E}^{(2)}]) 
Euler~[E_{infinity}] = \eta[{\cal E}^{(1)}].
\end{equation}

As said previously one must not be alarmed by seeing a
non-integer index value! Cantorian-fractal space-time
corresponds naturally to irrational numbers (irrational
conformal field theory). This result does in fact
correspond to the generalized Euler number associated with
the infinite-cycle intersection space $E_{infinity}$
(living inside the fractal string) and with a complex
dimension equal to $1/2$ the real dimension of the fractal
world sheet ${\cal E}^{(2)}$ given by $1+\phi$. This is
just the intrinsic dimension of a fractal bosonic random
walk. The fermionic dimension equals $1/2$ its value.

It is very important to emphasize that despite the fact
that the fractal dimension of the infinite-cycle
intersection space $E_{infinity}$ is $0$ this does {\bf
not} mean that such space is made of a point.
Cantorian-fractal space-time has no points. It corresponds
to a von Neumann non-commutative {\bf pointless } geometry
with a natural $p$-adic topology: Every point is the center
of a disc because the center of a disc can occupy many
different places simultaneously. This is the key to
understanding the wave-particle duality properties of an
indivisible quantum particle in the double-slit Young
experiment within the framework of negative probabilities
in Cantorian-fractal spaces \cite{nasccast},
\cite{bohm}.

Due to the ring structure of the golden mean, $\phi^n =
m\phi + n$ where $m,n$ are integers, Cantorian-fractal
space-time displays a Grassmanian nature. For a Grassmanian
number $\theta$ such that $\theta^2 = 0$ any function of
$\theta$ must be of the form $a\theta + b$. In a sense
Cantorian-fractal space-time encodes ``super-symmetry'' and
why the index formula is linked to a ``super-trace'': $n_+
- n_-$. The generalized Euler number of $E_{infinity}$ is
equal to $ 0 - \phi = -\phi$ meaning that the
Cantorian-fractal world sheet is left/right asymmetric. For
example, the Euler number associated with the triple
intersection of $S^1, S^0, S^{-1}$ was
\begin{equation}
-1 + \phi - \phi^2 = {\phi^3 \over 2} - (1 - {\phi^3 \over 2}) = \phi^3 -
1 = 2\phi - 2, etc. 
\end{equation}

Let's imagine that we want to generalize this result to
fractal $p$-branes. Imagine the volume of the $p$-brane is
given by the direct product ${\cal E}^{(n)}\times {\cal
E}^{(0)}$, where the first term represents the spatial
dimensions given by $(1+\phi)^{n-1}$ and the second one
corresponds to the discrete fractal flow of time
represented by dimension of ${\cal E}^{(0)}$ which is the
golden mean. The total dimension of the world volume is
given by the sum $(1+\phi)^{n-1}+\phi$.  The complex
dimension is one half that value. The spectral staircase
relation associated with the intersection of the following
cycles:
\begin{equation}
{\cal E}^{n} \wedge {\cal E}^{n-1} ... \wedge {\cal E}^{-s + 1} ...,
\label{eq:interseccn}
\end{equation}
is
\begin{eqnarray}
\eta[{\cal E}^{(n)}] &=& \sum_{dim >n} 1 - \sum_{dim <n} 1\nonumber\\
& = &-(n-1)+\zeta(0)=
\frac{1}{2}\left[(1+\phi)^{n-1}+\phi\right]\cdot\frac{(-1)^n\phi^{1-n}}{1+\phi}.
\end{eqnarray}

However such relation is {\bf not} always valid for an
arbitrary value of $n$!. In the last equation, the last
factor denotes the generalized Euler number of the
infinite-intersection cycle. We have seen that for fractal
strings, $n=1$, is valid. Are there other {\bf odd} values
of $n$ obeying such relation? A careful study shows that
{\bf the only} solution is the fractal string case $n=1$.

Now let's make a further generalization of the $p$-branes
case. The volume of the $p$-brane is given by the direct
product ${\cal E}^{(n)}\times {\cal E}^{(m)}$, where the
first term represents the spatial dimensions given by
$(1+\phi)^{n-1}$ and the second one corresponds to the
discrete fractal flow of time represented by dimension of
${\cal E}^{(m)}$ which is $(1+\phi)^{m-1}$. The total
dimension of the world volume is given by the sum
$(1+\phi)^{n-1}+(1+\phi)^{m-1}$. The complex dimension is
one half that value. The spectral staircase relation
associated with the intersection of the following cycles:

\begin{equation}
{\cal E}^{n+m} \wedge {\cal E}^{n+m-1} ... \wedge {\cal E}^{-s + 1} ...,
\label{eq:interseccnm}
\end{equation}
is
\begin{eqnarray}
\eta[{\cal E}^{(n+m)}] &=& \sum_{dim >n+m} 1 - \sum_{dim <n+m}
1 = -(n+m-1)+\zeta(0)\nonumber\\
&=&
\frac{1}{2}\left[(1+\phi)^{n-1}+(1+\phi)^{m-1}\right]\cdot\frac{(-1)^{n+m}\phi^{1-n-m}}{1+\phi}.
\end{eqnarray}

A careful analysis shows that the only solution, as before, is the fractal
string cases ($n=1$, $m=0$) or ($n=0$, $m=1$).

\section{\bf Concluding remarks}
\label{sec:conclu}

In section 2 we outlined the steps towards an heuristic
proof of the Riemann conjecture. The conjecture could be
proven heuristically, if and only if, a fractal/discrete
derivative operator is found obeying the requisites
outlined. The analog of the Riemann-Roch theorem in
Cantorian-fractal space-time was furnished. The index
formula for the derivative operator of zero ``conformal''
$U(1)$ weight, $restricted $ in the infinite-intersection
cycle, living inside the fractal world sheet, was found to
agree {\bf exactly} with the $\zeta (0) = - 1/2$. The
conformal weights were given in units of $1+\phi$ which
is the intrinsic fractal dimension of a bosonic random
walk. Fermionic random walks have $1/2$ the dimension of
the bosonic ones. The $\eta$ invariant which is related to
the spectral staircase associated with the spectral
dimensions of the infinite hierarchy of Cantor sets living
inside the (odd dimensional) fractal string, required a
zeta function regularisation $\sum 1 = \zeta (0) = - 1/2$.
The flow of time was discrete and fractal and corresponded
to the extra dimension of $\phi$ yielding a world sheet
${\cal E}^{(2)}$ of fractal dimension $1+\phi$.

Next, we entertained the opposite idea. And was found that
$\zeta$ {\bf cannot} have zeroes at
\begin{equation}
(1\pm i\nu)(\phi \pm i\nu) (\phi^2 \pm i\nu),\quad (1\pm
i\nu)({1\over\phi}\pm i\nu)({1\over\phi^2}\pm i\nu).\quad \nu =
\{\Im(s)~|\zeta(s) = 0\}. 
\end{equation}

Nevertheless, it is warranted to check if zeroes at $\phi +
i \nu$ and $(1- \phi) + i \nu$ exist. These points lie
inside the critical strip, $0< \Re(s) < 1$ and are {\bf
symmetrically} distributed with respect to the critical
line $Re(s) = 1/2$; i.e iff $s$ is a non-trivial zero then
$1-s$ must be as well as a direct consequence of the
functional equation obeyed by the zeta function inside the
critical strip (no poles in the gamma function). Since
$\phi + \phi^2 = 1$, it is plausible that there could be
zeroes of zeta at some values along the vertical lines
beginning at those points. which values of the imaginary
part? This is the question ... We checked some values but
not all of them. Perhaps the imaginary values do not
correspond to the imaginary values of $s = 1/2 + i\nu$ but
to some other unknown ones ...

Fractal $p$-adic strings were discussed by one of us in
\cite{cccastro}. The scattering of a particle off a $p$-adic
fractal string is another way to look at the $p$-adic
stochastic motion modelled by a SUSY QM system, whose
operator yields the imaginary parts of the zeroes of zeta.
It was suggested in \cite{cccastro} to establish the
correspondence, if any, between the exact location of the
poles of the scattering amplitudes of $p$-adic fractal
strings and the zeroes of the zeta.

The scattering amplitudes are given by a generalization of
the Veneziano formula in terms of the Euler gamma
functions. In the same fashion that the trivial poles of
the gamma functions in Eq. [\ref{eq:eulergamma}] yield the
trivial zeroes of zeta at $s=-2n$, $n=1,2,3,...$, it is
sensible to see if a similar correspondence can be
established, after an analytical continuation to the
critical strip $0\le s\le1$ is performed. Because one
expects complex-dimensions to play a fundamental role
\cite{lapidus} the question one may ask is if a one-to-one
correspondence between the nontrivial zeroes of zeta and
the spectrum (poles in the scattering amplitudes) of $p$-adic
fractal strings and Regge trajectories in the complex
angular momentum plane. In \cite{cccastro} we explicitly
defined the SUSY QM problem in terms of the two
iso-spectral Hamiltonians $H_+$, $H_-$ (self-adjoint) whose
eigenvalues where
\begin{equation}
\lambda_0=0,\quad\lambda_n=\lambda_n^+=\lambda_n^-,
\end{equation}
the imaginary parts of the zeroes of zeta.

The ``fused'' operator
\begin{equation}{\cal H}=\frac{1}{2}(H_+H_-+H_-H_+)+\frac{1}{4}
\end{equation}
was self-adjoint whose eigenvalues and eigenfunctions were
given by
\begin{equation}
{\cal H}\psi_n=(\frac{1}{2}+i\lambda_n)(\frac{1}{2}-i\lambda_n)\psi_n
=(\frac{1}{4}+\lambda_n^2) \psi_n,
\end{equation}
with
\begin{equation}
\psi_n={\cal F}^{-1}[{\cal F}(\psi_n^+)*{\cal F}(\psi_n^-)].
\end{equation}
${\cal F}$ and ${\cal F}^{-1}$ are inverse Fourier
transform pair. $\psi_n$ is related to the convolution star
product of the eigenfunctions $\psi_{n-1}^+$, $\psi_n^-$ of
the $H_+$, $H_-$ operators (even negative Witten parity)
and $n=1,2, ...\infty$.

It was argued in \cite{cccastro} that the $1/4$ coefficient
was intrinsically related to the superconformal fusion
rules properties of the eigenfunctions $\psi_{n-1}^+$,
$\psi_n^-$. Irreducible unitary highest weight
representations of the super-Virasoro algebra could select
and fix uniquely the value of the $1/4$ coefficient will be
an elegant proof of the Riemann conjecture since this
coefficient is due to the
$(1/2+i\lambda_n)(1/2-i\lambda_n)$ product of the zeroes
times their compex conjugates.

Pitk\"anen \cite{pitkanen} has argued that non-Hermitean
operators of the type $L_0\pm i V$ could yield the $1/2+
i\lambda_n$ zeroes of the zeta as an eigenvalue problem.
$L_0$ is the zero-mode of the Virasoro generator, or
dilatations in the complex plane. The potential $V$ is
linked to the $\lambda_n$ eigenvalues. Superconformal
invariance imposes the conditions $x$ = $n/2$ =
{\it half-integer-weights\/} like those appearing in the
Ramond fermionic string. Since $n=0,2$ are ruled out by the
Hadamard-Valle\`e de la Poussin theorem \cite{pitkanen}
this leaves $n=1$ as the only permitted value and hence
$x=1/2$.

If this approach is valid this entails that the propagator
of the $p$-adic fractal strings is nothing but the inverse
operator
\begin{equation}
[(L_0+i V)(L_0-i V)]^{-1}.
\end{equation}

Hence the poles of the propagator naturally correspond to
the zeroes of zeta. All this remains to be verified
explicitly. On the other hand we argued \cite{cccastro}
that the fused operator ${\cal H}$ whose eigenvalues were
$1/4+\lambda_n^2$ was quartic in derivatives as they should
be. The Schild action for a string is the square of the
Poisson brackets with respect to the world sheet variables
$(\sigma^0,\sigma^1)$. The momentum variable conjugate to
the area-variables of the string world sheet is called the
area-momentum. The square of such area-momentum is
precisely the Schild action. The analogy of the on-shell
mass condition for a point like particle $p_\mu^2+m^2=0$ is
\begin{equation}
P_{\mu\nu}^2+T^2=0,
\end{equation}
where $P_{\mu\nu}$ is the area-momentum of the string world
sheet, given by the Poisson bracket
\begin{equation}
P_{\mu\nu}=\{X_\mu,X_\nu\}_{\sigma_0,\sigma_1},
\end{equation}
with respect to the variables $\sigma_0,\sigma_1$ (world
sheet), with $X_\mu$ the string embedding coordinate in a
target spacetime of $D$ dimensions ($\mu$, $\nu$ = 1,2,
...D). $T$ is the string tension which has units of energy per
unit length.

The propagator is essentially the inverse of the operator
\begin{equation}
[P_{\mu\nu}^2+T^2]^{-1}.
\end{equation}
We are not including the center of mass degrees of freedom,
which should be incorporated. Hence in this Schild string
action model we can understand clearly why there is a
quartic-derivative operator and accordingly how the poles
of the scattering amplitudes, poles in the propagator,
could be related to the zeroes of zeta once we include
fractal chaos effects into the picture. The $p$-adic fractal
string is linked to a $p$-adic stochastic dynamics with a
Parisi-Sourles supersymmetry. Hence, the SUSY potential
$\Phi$ term must be incorporated into the momentum leaving
then $p\to p+\Phi$, which is just the usual minimal
coupling of a particle to an electromagnetic potential. The
inclusion of $\Phi$ mandated by the SUSY QM model
\cite{cccastro} into string theory is the correct way to
formulate the scattering amplitudes of the $p$-adic fractal
strings. Hence, the poles of their scattering amplitudes
could be linked to the zeroes of zeta or $1/4+\lambda_n^2$.
The inverse operator of the fused quartic derivative
Hamiltonian (the propagator) is
\begin{equation}
{\cal H}^{-1}=[\frac{1}{2}(H_+H_-+H_-H_+)+\frac{1}{4}]^{-1},
\end{equation}
and will have for poles the values of $1/4+\lambda_n^2$.
The $1/4$ coefficient can be interpreted as the zero-point
energy of the spectrum of the $p$-adic fractal string; in the
same way that $1/2$ is the zero-point energy of the
harmonic oscillator. Representation theory and super
conformal invariance could fix the value $1/4$ uniquely and
then we could have a proof of the Riemann conjecture.

Figures \ref{fig:hundred} and \ref{fig:thousand} show very
suggestive relations of the golden mean and the
distribution of the imaginary parts of the zeroes of zeta.
This could be related to the multifractal character of the
prime numbers distribution (See \cite{wolf}). Also other
relations of the Riemann's zeta function to fractal string
and M theory can be observed (See \cite{acj}).

\section*{\bf Acknowledgements} 

We acknowledge J\"orn Steuding, Johann Wolfgang
Goethe-Universit\"at Frankfurt, for giving us the reference
\cite{titchmarsh}. Also very constructive comments from
Matthew Watkins, School of Mathematical Sciences University
of Exeter, and Andrew Odlyzko's from ATT labs are
gratefully acknowledged. CC is indebted to E. Spallucci for
his kind hospitality in Trieste were this work was
completed, to A. Granik, J. Gonzalez, L. Reyes and R.
Guevara for many discussions. JMG is pleasured to
acknowledge the support of the present work by Centro de
Investigaciones en Ciencias Exactas y Naturales of the
Universidad de Antioquia, CIEN and the International Centre
for Theoretical Physics, ICTP.

\newpage

\section*{\bf Figures}
\label{sec:figs}

\def\baselinestretch{1}

\vfill
\newpage

\end{document}